\documentclass{article}
\usepackage{spconf,amsmath,graphicx}
\usepackage{amssymb}
\usepackage{bm}
\usepackage{xcolor}
\usepackage{qcircuit}
\usepackage{braket}
\usepackage{algorithm}
\usepackage{algorithmicx}
\usepackage{algpseudocode}
\usepackage{mathtools}
\usepackage{hyperref}
\usepackage{setspace}
\usepackage{microtype}

\renewcommand{\figureautorefname}{Figure~\negthinspace}

\renewcommand{\sectionautorefname}{Section~\negthinspace}

\title{Efficient quantum recurrent reinforcement learning via quantum reservoir computing}
\name{Samuel Yen-Chi Chen \thanks{The views expressed in this article are those of the authors and do not represent the views of Wells Fargo. This article is for informational purposes only. Nothing contained in this article should be construed as investment advice. Wells Fargo makes no express or implied warranties and expressly disclaims all legal, tax, and accounting implications related to this article.}}
\address{Wells Fargo}

\begin{document}
%
\maketitle
\begin{abstract}

Quantum reinforcement learning (QRL) has emerged as a framework to solve sequential decision-making tasks, showcasing empirical quantum advantages. A notable development is through quantum recurrent neural networks (QRNNs) for memory-intensive tasks such as partially observable environments. However, QRL models incorporating QRNN encounter challenges such as inefficient training of QRL with QRNN, given that the computation of gradients in QRNN is both computationally expensive and time-consuming. This work presents a novel approach to address this challenge by constructing QRL agents utilizing QRNN-based reservoirs, specifically employing quantum long short-term memory (QLSTM). QLSTM parameters are randomly initialized and fixed without training. The model is trained using the asynchronous advantage actor-aritic (A3C) algorithm. Through numerical simulations, we validate the efficacy of our QLSTM-Reservoir RL framework. Its performance is assessed on standard benchmarks, demonstrating comparable results to a fully trained QLSTM RL model with identical architecture and training settings.
\end{abstract}
\begin{keywords}
Quantum machine learning, Reinforcement learning, Recurrent neural networks, Long short-term memory, Reservoir computing
\end{keywords}
\section{Introduction}
\label{sec:intro}
\begin{figure}[htbp]
\centering
\includegraphics[width=1\linewidth]{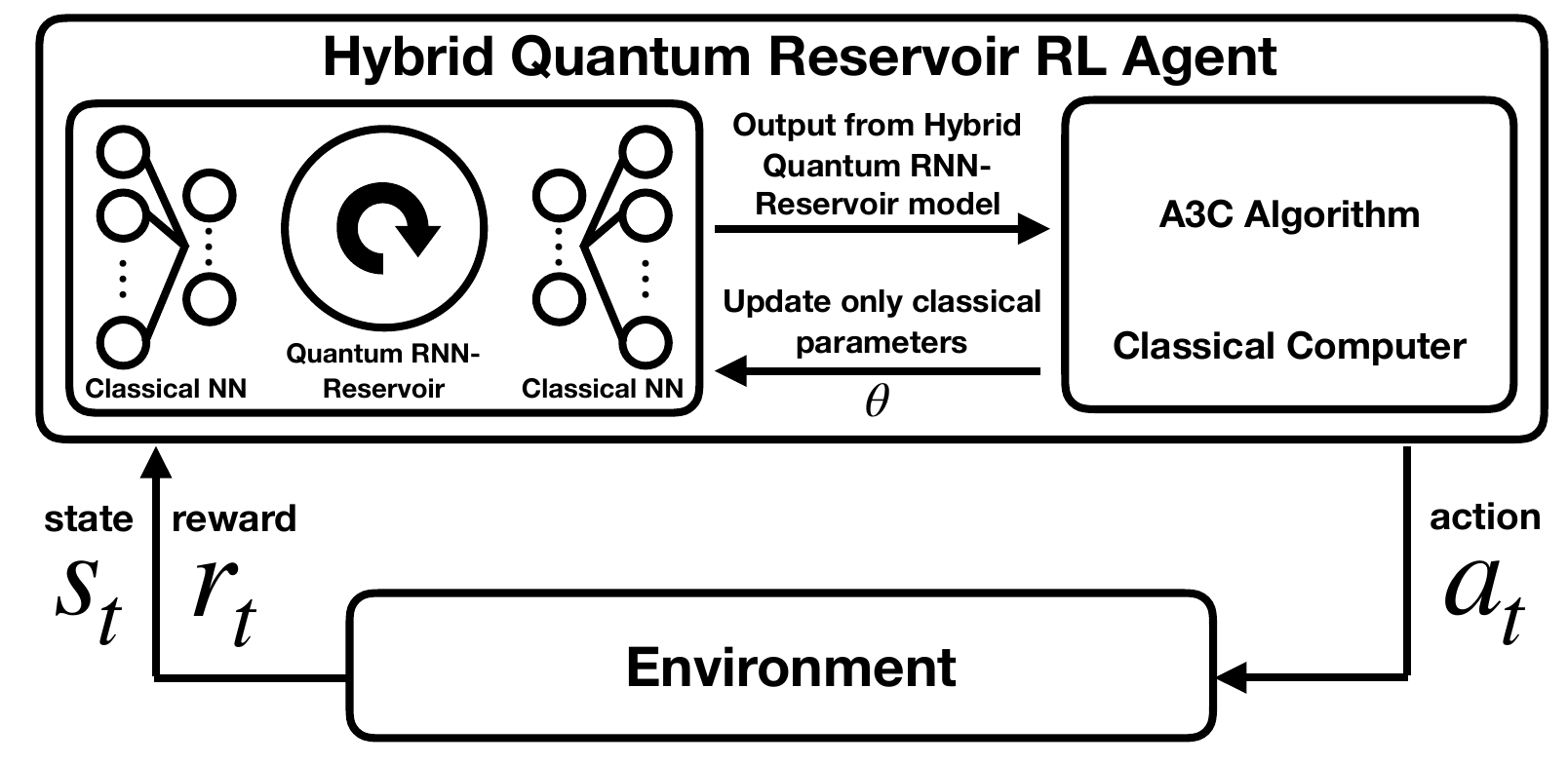}
\caption{{\bfseries The hybrid quantum-classical framework for QLSTM-RC-RL. }}
\label{fig:Overall}
\end{figure}

Quantum computing (QC) holds promise for enhanced performance in challenging computational tasks compared to classical counterparts. Yet, current quantum computers lack error correction, complicating deep quantum circuit implementation. These noisy intermediate-scale quantum (NISQ) devices \cite{preskill2018quantum} require specialized quantum circuit designs to fully exploit their potential advantages. A recent hybrid quantum-classical computing approach \cite{bharti2022noisy} leverages both realms, with quantum computers handling advantageous tasks while classical counterparts manage tasks like gradient calculations. Known as \emph{variational quantum algorithms}, these methods have excelled in specific machine learning (ML) tasks. Reinforcement learning (RL), a subset of ML concerned with sequential decision making, has achieved remarkable success through deep neural networks in complex tasks \cite{mnih2015human}. In contrast, the nascent field of quantum reinforcement learning (QRL) poses unexplored challenges. While most QRL approaches focus on variational quantum circuits (VQCs) without recurrence, a recent development introduces quantum recurrent neural networks (QRNNs) as seen in the work \cite{chen2023quantum}. This innovation demonstrates promise in partially observable environments, outperforming classical models. The challenge of training QRNNs is their computationally expensive gradient calculation. To address this, we propose the QLSTM-RC-RL framework. By harnessing quantum long short-term memory (QLSTM) \cite{chen2020quantum} and reservoir computing (RC) \cite{lukovsevivcius2009reservoir}, we optimize QRL training. QLSTM operates as an untrained \emph{reservoir}, with only the classical neural components before and after it undergoing training. The scheme is illustrated in \figureautorefname{\ref{fig:Overall}}. We further accelerate the training through the use of asynchronous training developed in the work \cite{mnih2016asynchronous,chen2023asynchronous}. Our numerical simulation shows that the proposed framework can reach performance comparable to fully trained counterparts when the model sizes are the same and under the same training setting. 
\section{Related Work}
\label{sec:related_work}
Quantum Reinforcement Learning (QRL) traces its origins to Dong et al.'s work \cite{dong2008quantum}. While traditionally requiring a quantum environment, recent VQC-based QRL advancements tackle classical settings. Chen et al. \cite{chen19} initiated this, addressing discrete environments like Frozen-Lake and Cognitive-Radio. Later, Lockwood et al. \cite{lockwood2020reinforcement} and Skolik et al. \cite{skolik2021quantum} expanded to continuous observation spaces, such as Cart-Pole. Chen et al. also introduced Quantum RNNs \cite{chen2023quantum} for partially observable scenarios, broadening QRL's practicality in classical contexts. In addition to learning value functions like $Q(s,a)$, recent QRL advances include policy function learning. Jerbi et al. \cite{jerbi2021variational} introduce quantum policy gradient RL using REINFORCE \cite{sutton2018reinforcement}. Hsiao et al. \cite{hsiao2022unentangled} enhance this with PPO and VQCs, showing quantum models with fewer parameters can surpass classical ones. This trend extends to various modified quantum policy gradient algorithms, including actor-critic \cite{schenk2022hybrid} and SAC \cite{lan2021variational}. QRL applies to quantum control, architecture search \cite{sequeira2022variational,10.1145/3588983.3596692} and multi-agent settings \cite{yun2022quantum,chao2023quantum}.
QRL optimization with evolutionary optimization is first studied in \cite{chen2022variational}. Asynchronous training of QRL is proposed in the work \cite{chen2023quantumasyncq,chen2023asynchronous}. Reservoir computing (RC) employing classical RNNs has undergone extensive research \cite{lukovsevivcius2009reservoir}, whereas the utilization of quantum RNNs such as QLSTM as reservoirs to perform time-series modeling represents a recent development \cite{chen2022reservoir}. The idea is to use QLSTM's internal dynamics and hidden states as a rich, dynamic memory or context for processing sequential data. This study integrates QRNN-based RL from \cite{chen2023quantum} with Quantum A3C from \cite{chen2023asynchronous}. Additionally, we demonstrate the effectiveness of using randomly initialized QRNNs, like QLSTM, as reservoirs, achieving comparable performance to fully-trained models and reducing training time.
\section{Reinforcement Learning}
\label{sec:RL}
\emph{Reinforcement learning} (RL) involves an agent interacting with an environment $\mathcal{E}$. At each time step $t$, the agent observes a \emph{state} $s_t$, selects an \emph{action} $a_t$ from the action space $\mathcal{A}$ according to its current \emph{policy} $\pi$, and receives a \emph{reward} $r_t$. The agent aims to maximize the expected return $V^\pi(s) = \mathbb{E}\left[R_t|s_t = s\right]$,  where $R_t = \sum_{t'=t}^{T} \gamma^{t'-t} r_{t'}$. It can also be defined using the action-value function $Q^\pi(s,a) = \mathbb{E}[R_t|s_t = s, a]$, which represents the expected return for taking action $a$ in state $s$ under policy $\pi$~\cite{sutton2018reinforcement}.
Unlike value-based methods (e.g., $Q$-learning), \emph{policy gradient} methods optimize $\pi(a|s;\theta)$ directly, updating parameters based on expected return, e.g., the REINFORCE algorithm~\cite{sutton2018reinforcement}.
In standard REINFORCE, $\theta$ updates via $\nabla_{\theta} \log \pi(a_t | s_t ; \theta) R_t$. High variance can be an issue. To reduce variance, subtract a learned state-dependent baseline, typically $V^\pi(s_t)$. This yields $\nabla_{\theta} \log \pi(a_t | s_t ; \theta)A(s_t, a_t)$, where $A(s_t, a_t) = Q(s_t, a_t) - V^\pi(s_t)$ is the \emph{advantage}. This method is called advantage actor-critic (A2C) with policy as the actor and the value function as the critic~\cite{sutton2018reinforcement}.
A3C (Asynchronous Advantage Actor-Critic) uses multiple concurrent actors for parallelized policy learning, improving stability and reducing memory needs. Diverse state encounters enhance numerical stability. A3C's efficient use of actors makes it a popular choice in reinforcement learning~\cite{mnih2016asynchronous} and has been studied in quantum RL recently \cite{chen2023asynchronous}.
\section{Variational Quantum Circuits}
\label{sec:VQC}
Variational quantum circuits (VQC), also known as parameterized quantum circuits (PQC) in the literature, are a distinctive class of quantum circuits containing trainable parameters. These parameters are optimized using classical machine learning methods, which can be gradient-based or gradient-free.
A VQC comprises three essential components. The \emph{encoding} block, denoted as $U(\mathbf{x})$, transforms classical data $\mathbf{x}$ into a quantum state. The \emph{variational} or \emph{parameterized} block, represented by $V(\boldsymbol{\theta})$, contains learnable parameters $\boldsymbol{\theta}$ optimized through gradient descent in this study. Finally, the \emph{measurement} phase outputs information by measuring a subset or all of the qubits, resulting in a classical bit string. Running the circuit once provides a bit string like "0,0,1,1." However, multiple circuit runs yield expectation values for each qubit. This paper specifically examines the Pauli-$Z$ expectation values from VQC measurements.
The mathematical expression of the VQC used in this work is $\overrightarrow{f(x ; \theta)}=\left(\left\langle\hat{Z}_1\right\rangle, \cdots,\left\langle\hat{Z}_N\right\rangle\right)$ , where $\left\langle\hat{Z}_{k}\right\rangle =\left\langle 0\left|U^{\dagger}(x)V^{\dagger}(\theta) \hat{Z_{k}} V(\theta)U(x)\right| 0\right\rangle$.

VQCs offer several advantages, including enhanced resilience to quantum device noise \cite{kandala2017hardware,mcclean2016theory}, which proves particularly valuable in the NISQ era \cite{preskill2018quantum}. Moreover, research has indicated that VQCs can exhibit greater expressiveness than classical neural networks \cite{sim2019expressibility,du2020expressive,abbas2021power} and can be trained effectively with smaller datasets \cite{caro2022generalization}. Noteworthy applications of VQC in QML span classification \cite{mitarai2018quantum,qi2023theoretical,chen2021end}, natural language processing \cite{yang2020decentralizing,yang2022bert,di2022dawn}, generative modeling \cite{chu2023iqgan} and sequence modeling \cite{chen2020quantum,bausch2020recurrent}.

\section{METHODS}
\label{sec:methods}
\subsection{QLSTM-Reservoir Computing}
\label{sec:qlstm}
\begin{figure}[htbp]
\centering
\includegraphics[width=1\linewidth]{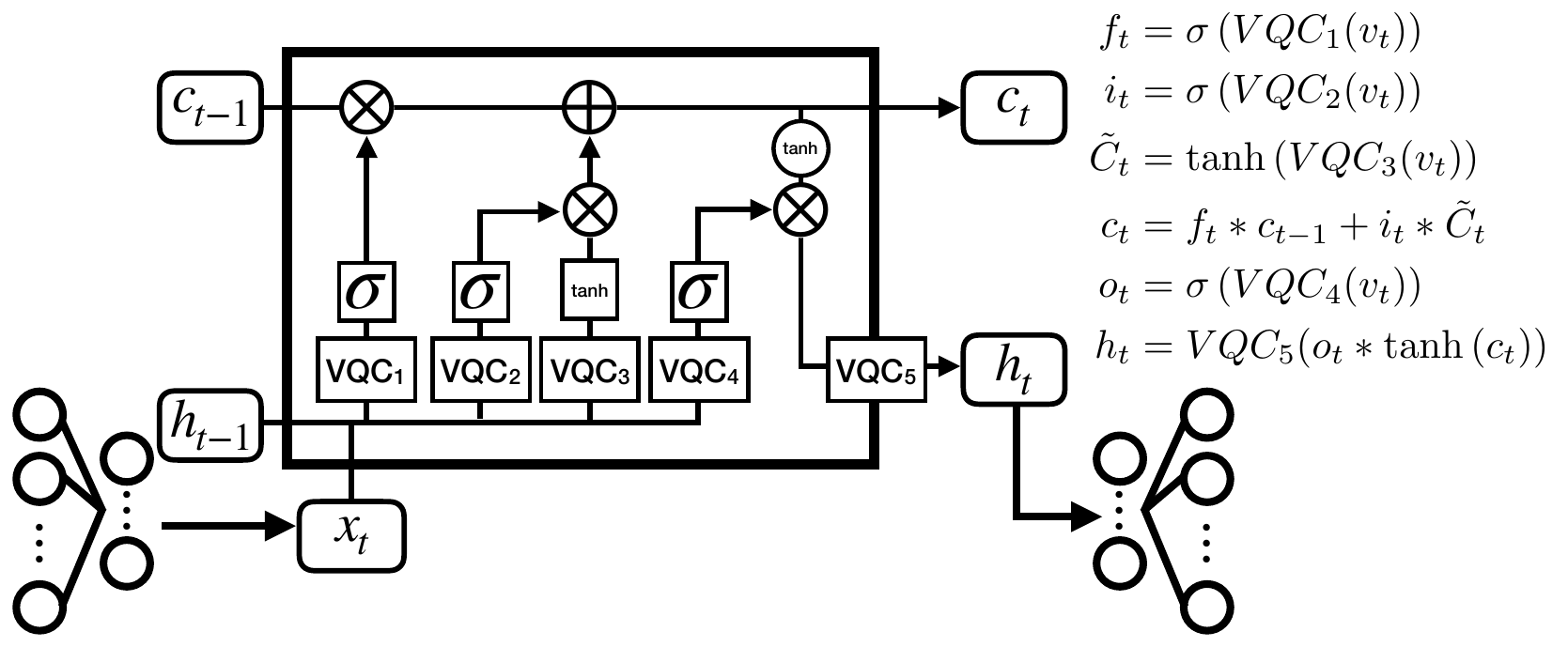}
\caption{{\bfseries The QLSTM-Reservoir used in the proposed QA3C.} QLSTM is first proposed in the work \cite{chen2020quantum}. In the proposed QLSTM-RC-RL framework, the VQC parameters are not trained. The input to the QLSTM is the concatenation $v_t$ of the hidden state $h_{t-1}$ from the previous time step and the current input vector $x_t$ which is the processed observation from the environment.}
\label{fig:QLSTM}
\end{figure}
\begin{figure}[htbp]
\begin{center}
\scalebox{0.55}{
\begin{minipage}{10cm}
\Qcircuit @C=1em @R=1em {
\lstick{\ket{0}} & \gate{H} & \gate{R_y(\arctan(x_1))} & \gate{R_z(\arctan(x_1^2))} & \ctrl{1}   & \qw       & \qw      & \targ    & \ctrl{2}   & \qw      & \targ    & \qw      & \gate{R(\alpha_1, \beta_1, \gamma_1)} & \meter \qw \\
\lstick{\ket{0}} & \gate{H} & \gate{R_y(\arctan(x_2))} & \gate{R_z(\arctan(x_2^2))} & \targ      & \ctrl{1}  & \qw      & \qw      & \qw        & \ctrl{2} & \qw      & \targ    & \gate{R(\alpha_2, \beta_2, \gamma_2)} & \meter \qw \\
\lstick{\ket{0}} & \gate{H} & \gate{R_y(\arctan(x_3))} & \gate{R_z(\arctan(x_3^2))} & \qw        & \targ     & \ctrl{1} & \qw      & \targ      & \qw      & \ctrl{-2}& \qw      & \gate{R(\alpha_3, \beta_3, \gamma_3)} & \meter \qw \\
\lstick{\ket{0}} & \gate{H} & \gate{R_y(\arctan(x_4))} & \gate{R_z(\arctan(x_4^2))} & \qw        & \qw       & \targ    & \ctrl{-3}& \qw        & \targ    & \qw      & \ctrl{-2}& \gate{R(\alpha_4, \beta_4, \gamma_4)} & \meter \gategroup{1}{5}{4}{13}{.7em}{--}\qw 
}
\end{minipage}}
\end{center}
\caption{{\bfseries VQC architecture for QLSTM.} The VQC architecture here is inspired by the work \cite{chen19}. The parameters $\alpha, \beta, \gamma$ are not trained in QLSTM-RC settings.}
\label{Fig:Basic_VQC_Hadamard_MoreEntangle}
\end{figure}
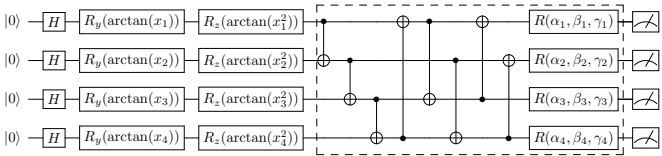

The QLSTM, depicted in \figureautorefname{\ref{fig:QLSTM}} and introduced by Chen et al. in \cite{chen2020quantum}, is a quantum adaptation of LSTM \cite{hochreiter1997long}. It employs VQCs instead of classical neural networks and excels in both time-series data and NLP tasks \cite{chen2020quantum,di2022dawn,li2023pqlm}.
The VQC incorporated into the QLSTM, as depicted in \figureautorefname{\ref{Fig:Basic_VQC_Hadamard_MoreEntangle}}, has shown impressive performance in time-series modeling \cite{chen2020quantum}. It encompasses data encoding via $R_{y}$ and $R_{z}$ rotations, a variational component involving CNOT gates for qubit entanglement, trainable unitary $R$ gates, and quantum measurement.
The original QLSTM proposal, applicable to time-series modeling \cite{chen2020quantum} and QRL \cite{chen2023quantum}, involves time-consuming training of VQC parameters. In this work, we adopt an approach using QLSTM as a reservoir to transform input data without the need for explicit VQC parameter training, as detailed in \cite{chen2022reservoir}. 
\subsection{QLSTM-RC-RL}
\label{sec:QLSTM_RC_RL}
The proposed QLSTM-RC-RL framework includes a \emph{dressed QLSTM model} consisting of classical neural networks for preprocessing and postprocessing, with a QLSTM in between. We adopted the quantum asynchronous advantage actor-critic (QA3C) training method developed in the work \cite{chen2023asynchronous}. Furthermore, we extend the original method to include the recurrent policy QLSTM. During asynchronous training, each local agent interacts with its own environment and stores the current trajectory in local memory. This trajectory is later used to calculate local gradients, which are then uploaded to the global shared model for updating.
\section{Experiments}
\label{sec:experiments}
\subsection{Environment}
\label{sec:environment}

\begin{figure}[htbp]
\centering
\includegraphics[width=1\linewidth]{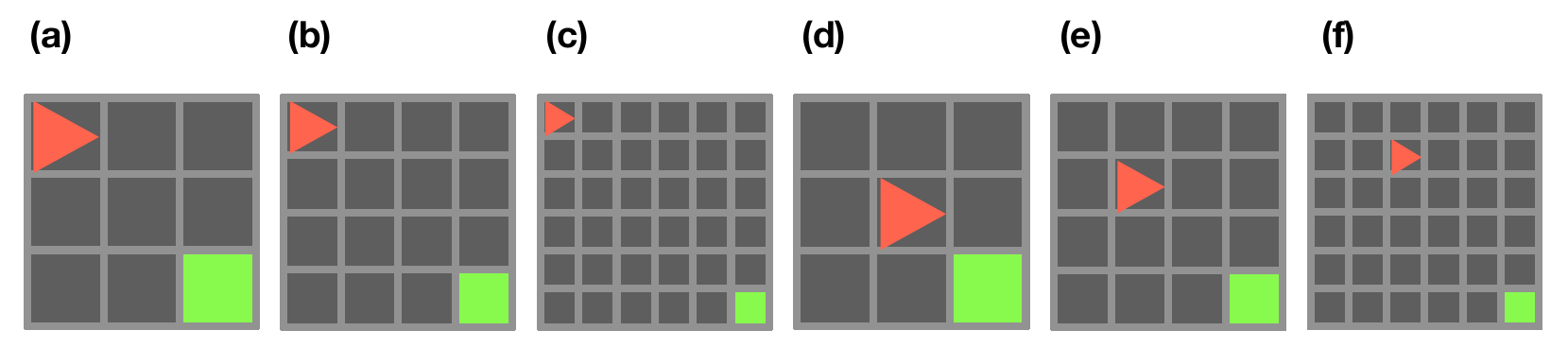}
\caption{{\bfseries The MiniGrid environments.} (a) - (c) are MiniGrid environments with fixed starting points and (d) - (f) are with random starting points (starting points shown in (d) - (f) are a set of examples). }
\label{fig:MiniGridEnv}
\end{figure}

In this study, we employ the MiniGrid-Empty environment, a widely utilized maze navigation scenario \cite{gym_minigrid}. The primary objective for our QRL agent is to effectively generate appropriate actions sequence based on the observations it receives at each time step, enabling it to traverse from the initial location to the designated destination, represented as the green box in \figureautorefname{\ref{fig:MiniGridEnv}}. Notably, the MiniGrid-Empty environment is characterized by a $147$-dimensional vector observation, denoted as $s_t$. It offers an action space $\mathcal{A}$ comprising six actions, namely \textit{turn left}, \textit{turn right}, \textit{move forward}, \textit{pick up an object}, \textit{drop the object being carried}, and \textit{toggle}. Of these actions, only the first three have practical consequences in this context, and the agent is expected to learn this distinction. Moreover, successful navigation to the goal rewards the agent with a score of $1$, albeit subject to a penalty determined by the formula $1 - 0.9 \times (\textit{number of steps}/\textit{max steps allowed})$, with the maximum allowable steps set at $4 \times n \times n$, where $n$ is the grid size \cite{gym_minigrid}. Throughout our experimentation, we explore various configurations, encompassing different grid sizes and variations in the initial starting points.
\subsection{Hyperparameters}
The hyperparameters for the proposed QLSTM-RC-RL are: Adam with learning rate: $1 \times 10^{-4}$, $beta_{1} = 0.92$, $beta_{2} = 0.999$, model lookup steps $L = 5$ and discount factor $\gamma = 0.9$.
The local agents/models calculate their own gradients every $L$ steps (the length of trajectory used during model updates) and update the model as described in \sectionautorefname{\ref{sec:QLSTM_RC_RL}}.
The number of parallel processes (local agents) is $80$.
\subsection{Model Size}
In our study, we employ hybrid QLSTM models composed of four key components: a classical NN for environmental observation preprocessing, a QLSTM that can be fully trained or initialized randomly and fixed in the RC scenario, and two classical NNs for processing QLSTM outputs to produce action logits and state values.
In our study, we utilize an $8$-qubit VQC-based QLSTM model with input and hidden dimensions of $4$. The internal state is $8$-dimensional. We explore QLSTM variations with $1$, $2$, and $4$ VQC layers, as shown in dashed box in \figureautorefname{\ref{Fig:Basic_VQC_Hadamard_MoreEntangle}}
All hybrid models share identical configurations for their classical neural networks. Specifically, the preprocessing NN consists of $147 \times 4 + 4 = 592$ parameters, the NN for action logits has $4 \times 6 + 6 = 30$ parameters, and the NN for state values comprises $4 \times 1 + 1 = 5$ parameters.
The number of parameters of QLSTM with $n$ VQC layer is: $8 \times 3 \times 5 \times n = 120n$ in which the VQCs are $8$-qubit and each general rotation rate is parameterized by $3$ parameters. There are $5$ VQCs in a QLSTM as shown in \figureautorefname{\ref{fig:QLSTM}}.
\subsection{Results}
\begin{figure}[htbp]
\includegraphics[width=1.\linewidth]{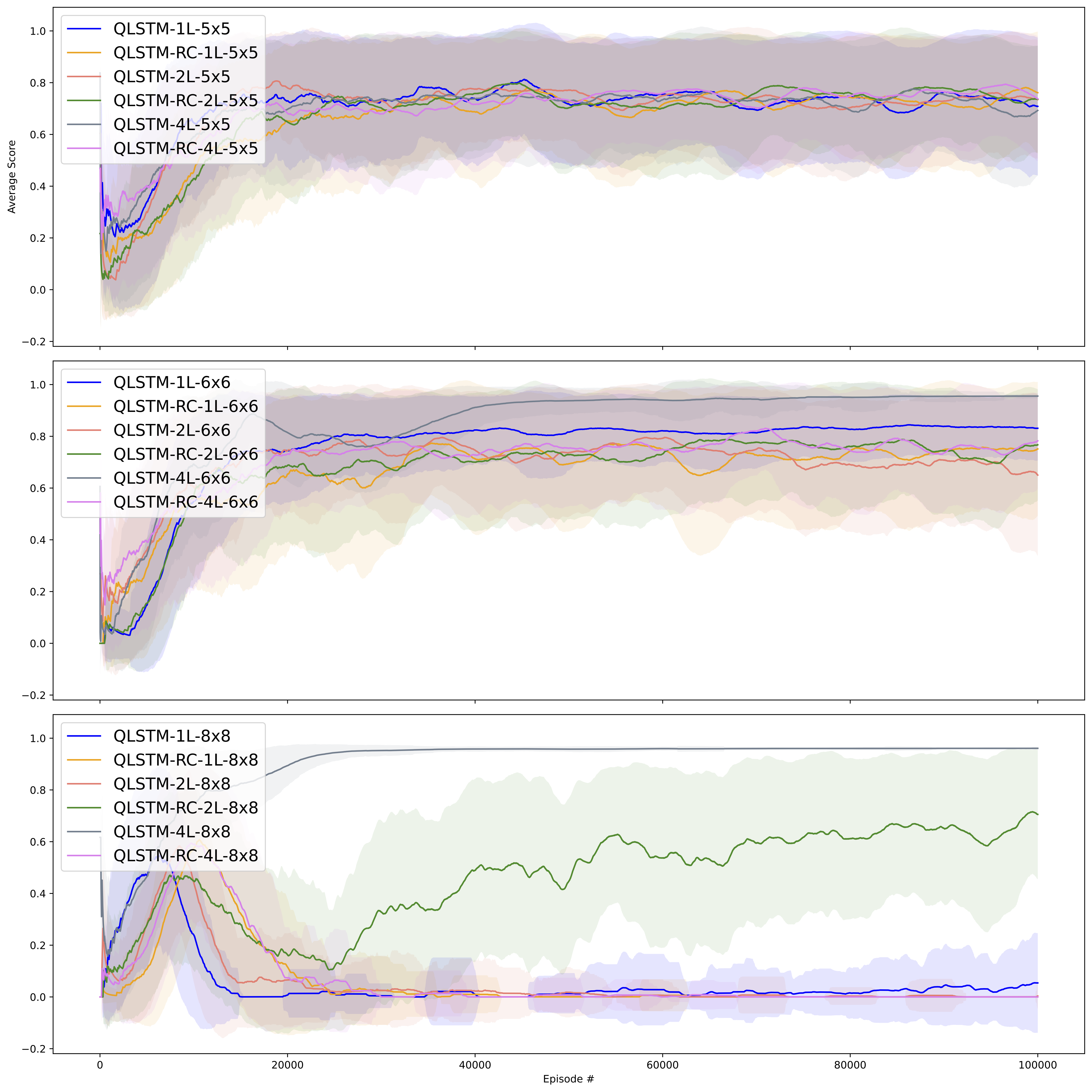}
\caption{{\bfseries Results: QLSTM-RC-RL in MiniGrid-Empty environment with fixed starting point.}}
\label{fig:results_non_random}
\end{figure}
\begin{figure}[htbp]
\includegraphics[width=1.\linewidth]{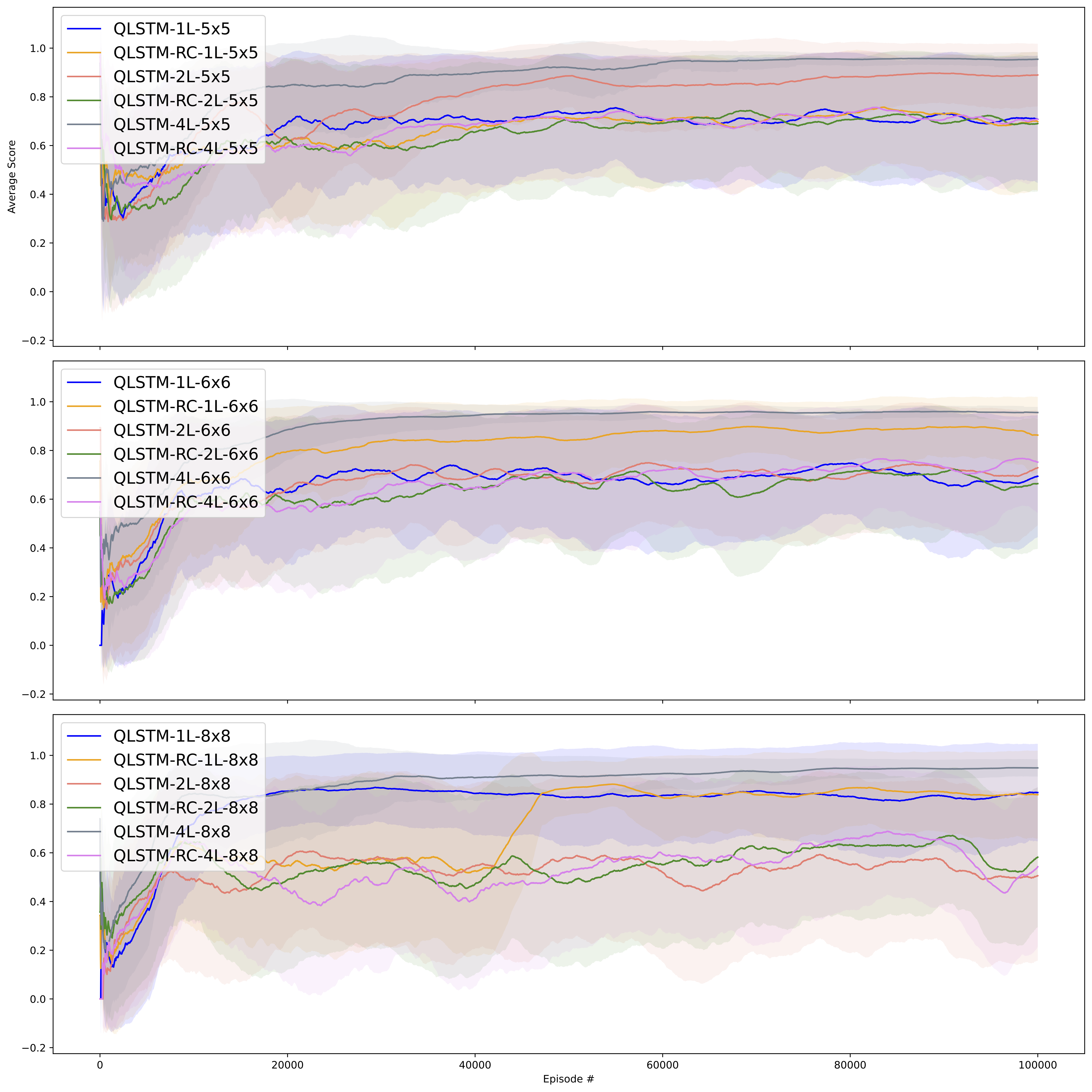}
\caption{{\bfseries Results: QLSTM-RC-RL in MiniGrid-Empty environment with random starting point.}}
\label{fig:results_random}
\end{figure}
\textbf{MiniGrid with fixed starting point} \quad We first consider the setting that the RL agent start from a fixed point in the environment. In the MiniGrid environment, this is at the upper-left corner of the maze as described in \sectionautorefname{\ref{sec:environment}} and \figureautorefname{\ref{fig:MiniGridEnv}}. 
The results are shown in the \figureautorefname{\ref{fig:results_non_random}}. 
We can observe that among the three environments settings we tested, QLSTM and QLSTM-RC with 1, 2 or 4 VQC layers reach similar performance in the MiniGrid-5x5. In the MiniGrid-6x6, fully-trained QLSTM with 4 VQC layers achieve the best performance. Other models still achieve good performance, which are close to the best one. The QLSTM-RC performs very similar to the fully-trained QLSTM. In the most difficult MiniGrid-8x8 case, only the fully-trained QLSTM with 4 VQC layers reaches the optimal performance. The QLSTM-RC with 2 VQC layers still learns but slowly. Other model configurations struggle to learn the good policies.\\
\textbf{MiniGrid with random starting point} \quad We further consider the setting that, in each episode, the RL agent starts from a random point in the environment. We provide a set of examples in \figureautorefname{\ref{fig:MiniGridEnv}. The results are shown in the \figureautorefname{\ref{fig:results_random}}.
We can observe that among the three environments settings we tested, the fully-trained QLSTM with 4 VQC layers outperform other models in all three cases. An interesting result is that the QLSTM-RC with only one VQC layer still reaches performance very close to the best performing agent. Other agents, either QLSTM or QLSTM-RC, perform very similar.
Overall, the performance of QLSTM and QLSTM-RC agents in this environment are comparable or superior than in the environments with fixed starting point. A possible reason is that the agent can experience different situations more frequently. The agent may start from a location closer to the goal and achieve the goal with a positive reward. This is crucial since the MiniGrid is a sparse environment and the agent may require a large number of trial and error to obtain a positive reward in the non-random environment.
\section{Conclusions}
\label{sec:conclusions}
In this paper, we first show the quantum recurrent neural network (QRNN)-based reservoir computing for RL. Specifically, we employ the hybrid QLSTM reservoir as the function approximator to realize the quantum A3C. From the results obtained from the testing environments we consider, our proposed framework shows stability and average scores comparable to their fully-trained counterparts in most testing cases when the model sizes, model architectures and training hyperparameters are fixed. The proposed method paves a new way of pursuing QRL with recurrence more efficiently.

\clearpage
\begin{spacing}{0.6}
\footnotesize
\bibliographystyle{IEEEbib}
\bibliography{bib/qml_examples,bib/rl,bib/tools,bib/qc_basic,bib/vqc,bib/qrl,bib/classical_ml,bib/rc}
\end{spacing}
\end{document}